\documentclass[
    prd, twocolumn, superscriptaddress, nofootinbib, amsmath, amssymb,
    aps, floatfix, preprintnumbers
]{revtex4-2}
\usepackage[utf8]{inputenc}
\usepackage{graphicx,xcolor,setspace}
\usepackage[
    colorlinks=true,
    linkcolor=blue,
    urlcolor=blue,
    citecolor=blue
]{hyperref}
\usepackage{multirow}
\usepackage{braket}
\usepackage{physics}

\usepackage{bm}

\def\beq{\begin{equation}}
\def\eeq{\end{equation}}
\def\bea{\begin{eqnarray}}
\def\eea{\end{eqnarray}}

\usepackage{verbatim}

\begin{document}
\title{Low-Energy Compton Scattering in Materials}
\author{Rouven Essig}
\affiliation{C.N. Yang Institute for Theoretical Physics, Stony Brook University, NY 11794, USA}
\author{Yonit Hochberg}
\affiliation{Racah Institute of Physics, Hebrew University of Jerusalem, Jerusalem 91904, Israel}
\author{Yutaro Shoji}
\affiliation{Racah Institute of Physics, Hebrew University of Jerusalem, Jerusalem 91904, Israel}
\author{Aman Singal}
\affiliation{C.N. Yang Institute for Theoretical Physics, Stony Brook University, NY 11794, USA}
\author{Gregory Suczewski}
\affiliation{C.N. Yang Institute for Theoretical Physics, Stony Brook University, NY 11794, USA}

\begin{abstract}
Low-energy Compton scattering is an important background for sub-GeV dark matter direct-detection and other experiments. Current Compton scattering calculations typically rely on assumptions that are not valid in the low-energy region of interest, beneath $\sim 50$~eV. Here we relate the low-energy Compton scattering differential cross section to the dielectric response of the material. Our new approach can be used for a wide range of materials and includes all-electron, band structure, and collective effects, which can be particularly relevant at low energies. We demonstrate the strength of our approach in several solid-state systems, in particular Si, Ge, GaAs, and SiC, which are relevant for current and proposed experiments searching for dark matter, neutrinos, and millicharged particles. 
\end{abstract}

\preprint{YITP-SB-2023-29}

\maketitle

\section{Introduction}

Tremendous progress has been made in dark matter~(DM) direct-detection experiments in the last decade. Although several direct-detection experiments are now searching for DM with sub-GeV masses through their interactions with electrons (see, {\it  e.g.}, Ref.~\cite{Essig:2022dfa} and references therein), this mass region remains to be  fully explored. A crucial ingredient of direct-detection searches for DM is a detailed understanding of backgrounds. It is especially important to understand backgrounds that lie in the $\sim$1--50~eV energy range, since the relevant energy transfer in a collision between a sub-GeV DM particle and electrons in, {\it e.g.},  semiconductors is typically a few eV, decaying rapidly for higher energies~\cite{Essig:2011nj}. 

Environmental photons that Compton scatter off detector electrons can produce low-energy ionization events and thus constitute an important background in experiments searching for sub-GeV DM.  It is therefore imperative to have reliable calculations of the Compton scattering cross section and spectrum down to the lowest energies probed by an experiment. The differential Compton scattering cross section at low-energies is most commonly calculated using the Relativistic Impulse Approximation~(RIA). The RIA is implemented in many computational softwares, including {\tt Geant4}~\cite{GEANT4:2002zbu,Allison:2006ve,Allison:2016lfl}. The \texttt{FEFF} program~\cite{Rehr:2000FEFF,Rehr:2009FEFF, Rehr:2010FEFF, Kas:2021} improves on the RIA and was found in Ref.~\cite{DAMIC-M:2022xtp} to agree better than RIA with data at energies $\gtrsim 100$~eV. However, RIA does not incorporate band-structure effects, while \texttt{FEFF} includes these effects only through the Korringa-Kohn-Rostoker Green function method~\cite{Korringa:1947KKR, Kohn:1954KKR} and a `muffin-tin' potential.  Moreover, both RIA and \texttt{FEFF} assume the photon interacts with a single electron and neglect collective effects of the electron density.  These assumptions are typically not valid at energies below $\order{50~\textrm{eV}}$ (depending on the material) where only valence electrons can be excited. 

In this work, we develop a new approach for calculating the Compton scattering cross section in the low-energy regime. In particular, we relate the differential Compton cross section to the response of the material via the dielectric function of the target. This allows us to relate the Compton rate to directly measurable properties of a detector. The dielectric function formalism contains the full material response, including the band structure and many-body effects, and  can be calculated using existing codes, analytical models, or, in principle, obtained from data. As we will see, the full response of the material is indeed important for low-energy Compton scattering and is crucial in at least some of the materials we consider. 

Our results will enable a better understanding of the expected Compton-scattering backgrounds in direct-detection experiments targeted at sub-GeV DM masses~\cite{Essig:2022dfa} and experiments searching for neutrino scattering at low energies~\cite{Fernandez-Moroni:2021nap}, and help distinguish them from potential beyond-the-Standard-Model signals. Our work can additionally be used to calibrate detectors for low-energy scattering events~\cite{Botti:2022lkm,DAMIC-M:2022xtp} and to probe our understanding of scattering with electrons bound in semiconductors. 

This paper is organized as follows.  Section~\ref{sec:RIA} reviews the RIA method that is most commonly used for computing low-energy Compton scattering. In Section~\ref{sec:dielectric}, we present our new approach for calculating the Compton scattering differential cross section in terms of the dielectric function, enhancing accuracy at low energies. Section~\ref{sec:results} presents our results for Compton scattering rates in various materials, with particular emphasis on silicon. We summarize in Section~\ref{sec:summary}. A set of Appendices elaborate on single electron excitations, details of the density functional theory we use, and why the Lindhard model works well for our purposes.

\section{Relativistic Impulse Approximation (RIA)}\label{sec:RIA}

The spin-averaged differential cross section for free-electron Compton scattering in the laboratory frame is well known~\cite{Peskin:1995ev},
\begin{align}\label{kleinnishina}
	\frac{d \sigma}{d \cos \theta} = \frac{\pi \alpha^{2}}{m^{2}} \left( \frac{\omega'}{\omega}\right)^{2} \left(\frac{\omega'}{\omega} + \frac{\omega}{\omega'} - \sin^{2} \theta\right), 
\end{align}
where $\omega$ is the initial photon energy, $\omega'$ is the final photon energy, $m$ is the mass of the electron, $\alpha$ is the electromagnetic fine-structure constant, and $\theta$ is the angle between the initial and final photon directions. For electrons that are not free, the above Compton scattering cross section must be modified. The most common method of calculation is known as the Relativistic Impulse Approximation~(RIA), which assumes the photon scatters off of a single electron with a momentum distribution. 

In the RIA formulation, the differential cross section can be factored into two pieces, one pertaining to the free-electron-photon interaction and the other, known as the Compton profile, capturing the many-electron effects (typically only for an isolated atom) via the momentum distribution of the electron~\cite{Ribberfors_1975_1, Ribberfors_1975_2, Brusa_1996}. It is convenient to define the momentum transfer vector, $\vec{q} = \vec{k'} - \vec{k} = \vec{p} - \vec{p'}$, where $\vec{k}$ ($\vec{p}$) and $\vec{k'}$ ($\vec{p'}$) are the initial and final momenta of the electron (photon) in the laboratory frame. The energy of the electron is $E\left(k\right) = \sqrt{k^{2} + m^{2}}$. Then, the differential cross section per unit cell is given by 
\begin{align}\label{RIA}
	\left(\frac{d^{2} \sigma}{d \omega' d \Omega'}\right)_{\rm RIA} = N_{\rm atoms} \frac{\alpha^2}{2 q E\left(k_{z}\right)} \frac{\omega'}{\omega}\overline{X}\left(k_{z}\right) J\left(k_{z}\right)\ .
\end{align} 
Here, we have assumed the crystal is composed of a single element, and $N_{\rm atoms}$ is the number of atoms per cell. If there are multiple elements, the cross section is given by the sum of the contributions of each. 

The Compton profile, $J\left(k_{z}\right)$, is defined as 
\begin{align}\label{Compton profile}
    J\left(k_{z}\right) = \int \rho(\vec{k}) dk_{x} d k_{y}\ ,
\end{align} 
where $\rho(\vec{k})$ is the ground-state electron-momentum density. The kernel function, $\overline{X}$, is proportional to the squared-averaged amplitude for free-electron Compton scattering, 
\begin{align}\label{}
	\overline{X}\left(k_{z}\right) &= \frac{K_{f}\left(k_{z}\right)}{K_{i}\left(k_{z}\right)} +\frac{K_{i}\left(k_{z}\right)}{K_{f}\left(k_{z}\right)}\nonumber \\
								&\hspace{3ex}+2m^{2}\left(\frac{1}{K_{i}\left(k_{z}\right)}-\frac{1}{K_{f}\left(k_{z}\right)}\right)\nonumber\\
								&\hspace{3ex}+m^{4} \left(\frac{1}{K_{i}\left(k_{z}\right)}-\frac{1}{K_{f}\left(k_{z}\right)}\right)^{2}\ ,
\end{align} 
evaluated at 
\begin{align}\label{}
	k_{z} &= -\frac{\vec{k}\cdot  \vec{q}}{q}\nonumber\\
 &= \frac{\omega \omega' \left(1-\cos \theta\right) - E\left(k_{z}\right)\left(\omega - \omega'\right)}{q}\ ,
\end{align}
where 
\begin{align}\label{}
    K_{i}\left(k_{z}\right) = E\left(k_{z}\right) \omega + \frac{\omega \left(\omega - \omega' \cos \theta\right)k_{z}}{q}
\end{align} 
and
\begin{align}\label{}
    K_{f}\left(k_{z}\right) = K_{i}\left(k_{z}\right) - \omega \omega' \left(1- \cos \theta\right)\ .
\end{align}

To include the effect of the binding energy, we consider only the energetically allowed electrons in the Compton profile,
\begin{equation}
    \bar J(p_z,\Delta E)=\sum_i\Theta(\Delta E-E_i)\int \rho_i(\vec{k}) dk_{x} d k_{y}\ ,
\end{equation}
where $E_i$ and $\rho_i(\vec{k})$ are the ionization energy and the electron density of the $i$-th electron. In our numerical analysis, we use the Compton profiles listed in Ref.~\cite{BIGGS1975201} together with the atomic binding energies computed by {\tt cFAC}~\cite{doi:10.1139/p07-197}, which we list in Table~\ref{table_binding_energy}.

The approach described in this section is widely used for Compton scattering calculations for electron recoil energies of $\order{\text{keV}}$, but it becomes inaccurate for lower energies, as it fails to incorporate band-structure effects and the full response of the material.  We discuss how to include these effects in the next section. 

\begin{table}
\begin{center}
    \begin{tabular}{|c||ccc|}\hline
        Atom & \multicolumn{3}{c|}{Binding Energy (eV)} \\
        \hline\hline
        C & $9.76[2p^2]$ & $17.6[2s^2]$ & $-$ \\
        Si & $6.39[3p^2]$ & $13.2[3s^2]$ & $-$ \\
        Ga & $5.32[4p^1]$ & $12.1[4s^2]$ & $24.8[3d^{10}]$ \\
        Ge & $6.54[4p^2]$ & $14.8[4s^2]$ & $35.3[3d^{10}]$ \\
        As & $7.83[4p^3]$ & $17.6[4s^2]$ & $26.9[3d^{10}]$\\
        \hline
    \end{tabular}
    \caption{The binding energies of valence electrons for isolated atoms computed by {\tt cFAC}~\cite{doi:10.1139/p07-197} ($\Delta E<100~\rm{eV}$). The corresponding orbitals are indicated together.}
    \label{table_binding_energy}
\end{center}
\end{table}

\section{Dielectric response}\label{sec:dielectric}

    We now present our new approach to computing the differential Compton scattering cross section, utilizing the material response of any target material. 

    The interaction Hamiltonian describing the non-relativistic electron-photon interaction in the $A^{2}$ approximation is given by 
    \begin{equation}
        V_I = \frac{e^2}{2m} \sum_a |\vec{A}(\vec{x}_a,t)|^2\ .
    \end{equation}
    Using Fermi's Golden Rule, the differential cross section for Compton scattering in the material is 
    \begin{align}
          \frac{d\sigma_{i \to f}}{d\Omega' d\omega'}&=\frac{\omega'^{2}}{(2\pi)^3}|\bra{\vec{p}_f,\lambda_f,S_f} V_{I}\ket{\vec{p}_i,\lambda_i,S_i}|^{2}\nonumber\\
          &\times (2\pi)\delta(E_f+\omega'-E_i-\omega)\ .
    \end{align}
    $S_i$ and $S_f$ represent the initial and final states of the target material, and $\lambda_i$ and $\lambda_f$ represent the initial and final polarization states of the photon. The interaction matrix can be factored such that 
    \begin{align}\label{eq:base}
         \bra{\vec{p}_f,\lambda_f,S_f} & V_I\ket{\vec{p}_i,\lambda_i,S_i}=\frac{e^2}{2m \sqrt{\omega' \omega}}(\vec{\epsilon}_{\lambda_i, \vec{p}_i}\cdot\vec{\epsilon}_{\lambda_f, \vec{p}_f}^{\,\,*})\nonumber\\
         &\times\sum_a\bra{S_f}e^{i\left(\vec{p}_i - \vec{p}_f\right)\cdot\vec{x}}\ket{S_i}\ .
    \end{align}
    
    Averaging (summing) over the inital (final) photon polarizations, we can make the replacement $|\vec{\epsilon}_{\lambda_i, \vec{p}_i}\cdot\vec{\epsilon}_{\lambda_f, \vec{p}_f}^{\,\,*}|^{2} \to (1+\cos^{2}\theta)/2$. Summing over all final states of the material and averaging over the direction of the momentum transfers, we can write the differential cross section in the form
    \begin{align}\label{eq:diff}
          \frac{d\sigma_{\omega}}{d\Omega' d\omega' d(\Delta E)}=&\frac{ N_{\rm cell}\alpha^{2} \omega'}{2 m^{2} \omega} \left(1+\cos^{2} \theta \right)\nonumber\\ & \qquad\times\delta(\Delta E +\omega'-\omega)\frac{d f}{d (\Delta E)}\ , 
    \end{align}
    where
    \begin{align}\label{eq:collective_f}
        \frac{d f}{d(\Delta E)} = \frac{1}{N_{\rm cell}} &\sum_{f} \delta (E_{f} -E_{i} - \Delta E)\nonumber \\
        &\times \int \frac{d \Omega_{\hat{q}}}{4 \pi}|\bra{S_f}e^{i\vec{q}\cdot \vec{x}}\ket{S_i}|^{2} \ .
    \end{align}
    Here $\vec{q} = \hat{q}q = \vec{p}_i - \vec{p}_f$ is the momentum transferred from the incoming photon to the detector target material, $\Omega_{\hat{q}}$ is the solid angle in the direction of $\vec{q}$, and $N_\mathrm{cell}$ is the number of unit cells in the crystal. 
    
    The above can be directly related to the dynamic structure factor at zero temperature $S(\Delta E, \vec{q})$, defined as~\cite{Sturm1993} 
    \begin{align}
        S(\Delta E, \vec{q}) =\ &\frac{2\pi}{V}\sum_f \delta\left(E_f - E_i - \Delta E\right) \\&\qquad\qquad\quad\times|\bra{S_f}{e^{i\vec{q}\cdot\vec{x}}}\ket{S_i}|^{2}\ , \nonumber 
    \end{align}
    and so
    \begin{align}
        \frac{d f}{d(\Delta E)} = \frac{V_\mathrm{cell}}{2\pi}\int\frac{d \Omega_{\hat{q}}}{4\pi} S(\Delta E, \vec{q})\ .
    \end{align}
    Here, $V$ is the volume of the target material, and $V_\mathrm{cell}\equiv V/N_\mathrm{cell}$ is the volume of the unit cell. The structure factor is equivalently written in terms of the dielectric function $\epsilon(\Delta E, \vec{q})$ as \cite{Sturm1993}
         \begin{align}
        S(\Delta E, \vec{q}) =\ &\frac{q^2}{2\pi\alpha}\mathrm{Im}\left\{\frac{-1}{\epsilon(\Delta E, \vec{q})}\right\}\ , 
    \end{align}
    and so Eq.~\eqref{eq:collective_f} becomes
    \begin{align}
        \frac{d f}{d(\Delta E)} = \frac{q^2V_{\rm cell}}{4\pi^2\alpha}\mathrm{Im}\left\{\frac{-1}{\epsilon(\Delta E, q)}\right\}\ .
    \end{align}
    Here, $\epsilon(\Delta E, q)$ is defined as the directional average of $\epsilon(\Delta E, \vec{q})$.
    
    Rewriting the solid angle $\Omega^\prime$ in terms of the momentum transfer $q$ and integrating over the final photon energy $\omega^\prime$ in Eq.~\eqref{eq:diff}, we arrive at the differential Compton cross section per unit cell in terms of the dielectric function of the material, 
    \begin{align}\label{eq:res_die}
        \frac{d\sigma_\omega}{d(d\Delta E)dq} = \frac{\alpha V_{\rm cell}}{4\pi m^2} \frac{q^{3}}{\omega^{2}}(1+\cos^{2}\theta)\textrm{Im}\left(\frac{-1}{\epsilon(\Delta E, q)}\right)\ .
    \end{align}
    The above equation is the main result of this paper, which  we will use to compute the Compton scattering rates in various target materials relevant for sub-GeV DM searches. Below we will present our results using an analytical approximation for the dielectric function, as well as using numerical tools.

    The analytical approximation to the dielectric function that we use in this work is the Lindhard dielectric function \cite{Hochberg:2021pkt,Knapen:2021run}, given by
    \begin{align}\label{eq:Lind}
        \epsilon_{\rm Lind}
        =1+\frac{3\omega_p^2}{q^2v_F^2}\qty(\frac12+F_++F_-)\ ,
    \end{align}
    where
    \begin{align}
    F_\pm&=\frac{k_F}{4q}\qty(1-Q_\pm^2)\log^{\rm PV}\frac{Q_\pm+1}{Q_\pm-1},\\
        Q_\pm&=\frac{q}{2k_F}\pm\frac{m_e}{qk_F}(\omega+i\Gamma_p)\ .
    \end{align}
    Here,  $\omega_p=\sqrt{(4\pi\alpha n_e)/m_e}$ is the plasma frequency, and $k_F=(3\pi^2n_e)^{1/3}$ is the Fermi momentum. For the width of the plasmon peak, $\Gamma_p$, we adopt $10\%$ of the plasmon peak frequency.

    We also employ several numerical methods to compute the dielectric function. We use the GPAW calculation with Local Field Effects (LFEs) dielectric function from \texttt{DarkELF}~\cite{Knapen:2021run, Knapen:2021bwg}, and the \textit{valence--to--conduction} only dielectric function from \texttt{EXCEED-DM}~\cite{Griffin:2021exdm, Trickle:2022exdm}.
    
    \begin{figure*}
        \centering
        \includegraphics[width = 0.497\linewidth]{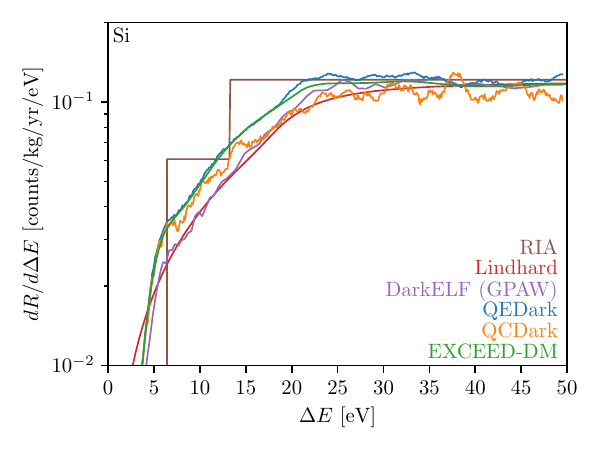}
        \includegraphics[width = 0.497\linewidth]{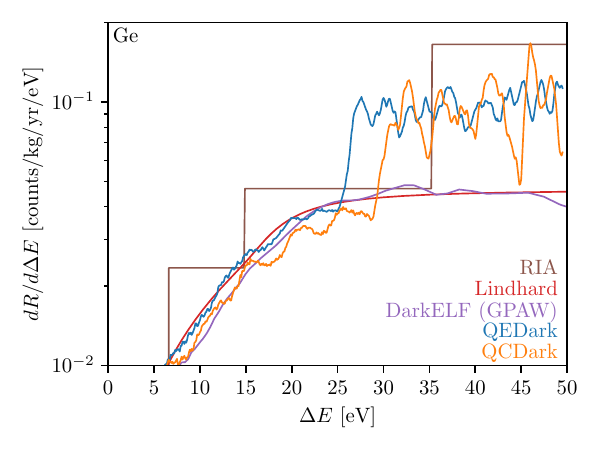}
        \includegraphics[width = 0.497\linewidth]{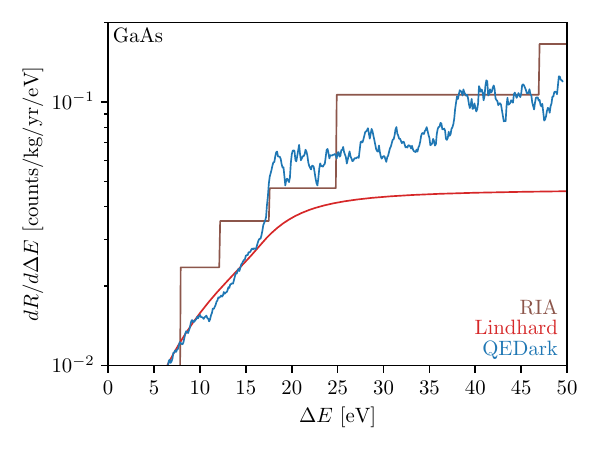}
        \includegraphics[width = 0.497\linewidth]{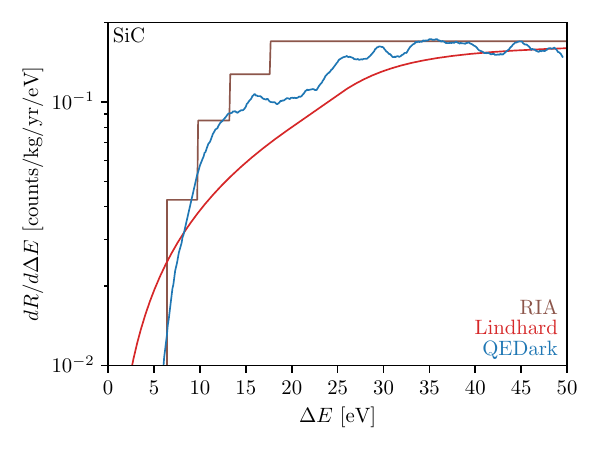}
        \caption{{\bf Compton scattering rates.} Compton scattering rates for incoming photons with $E_\gamma=1.461~\rm MeV$ and $n_\gamma=1.25\times10^{-14}~\rm cm^{-3}$. The {\it top-left} panel shows the rates for Si, {\it top-right} for Ge, {\it bottom-left} for GaAs, and {\it bottom-right} for SiC (3C) crystals. RIA is described in Section~\ref{sec:RIA} and the Lindhard model in Eq.~\eqref{eq:Lind}. For \texttt{DarkELF}, we use the GPAW results with local field effects from Ref.~\cite{Knapen:2021run, Knapen:2021bwg}. For \texttt{QEDark}~\cite{Essig:2015cda} and \texttt{QCDark}~\cite{Dreyer:2023ovn}, we use Eq.~\eqref{eq:dielectric_cff} to calculate the imaginary part of the dielectric function and use the real part of the Lindhard dielectric function when constructing the full material response. We use the available \textit{valence--to--conduction only} results from \texttt{EXCEED-DM}~\cite{Trickle:2022exdm}.}
        \label{fig:cs_rates}
    \end{figure*}
    \begin{figure}
        \centering
        \includegraphics[width=\linewidth]{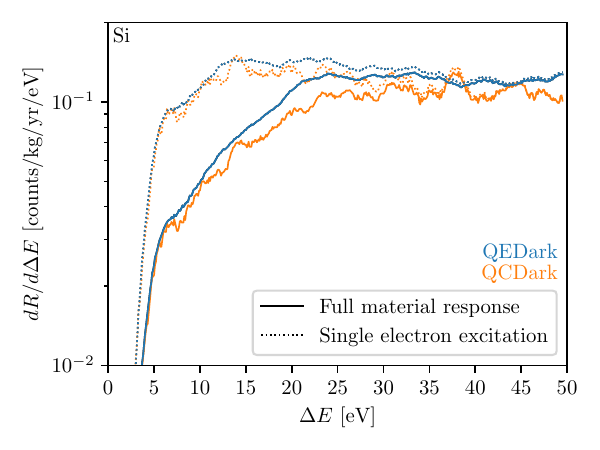}
        \caption{{\bf Material response.} 
        The effect of collective behavior of electrons in the context of Compton scattering rates in Si for incoming photons with $E_\gamma=1.461~\rm MeV$ and $n_\gamma=1.25\times10^{-14}~\rm cm^{-3}$. The dashed lines only include the single-electron excitations that are captured by the imaginary part of the dielectric function, while the solid lines include the full material response, which differs from the former by a factor of $1/\abs{\epsilon(\Delta E, q)}^2$. In constructing the full material response, we include the real part of $\epsilon(\Delta E, q)$ from the Lindhard dielectric function, Eq.~\eqref{eq:Lind}.}
        \label{fig:screening}
    \end{figure}
    
    Alternatively, we can employ a single electron excitation approximation, where the photon excites an electron from the occupied to unoccupied bands (see Appendix \ref{app:relativistic} for more details).
    However, in reality, the low energy excited states have very short lifetimes and de-excite by emitting low energy photons, which are then immediately absorbed by the material. The correct out states are thus dressed by the sum of such self-energy diagrams, which results in the factor of $1/\left|\epsilon(\Delta E, q)\right|^{2}$.
    Thus, the single electron excitation approximation generally misses this factor in the target material response, including for Compton scattering and DM--electron scattering. We will show the effects of omission of these collective effects in the context of Compton scattering in Fig.~\ref{fig:screening}. 

    The single electron excitation response is generally computed in terms of the crystal form factor~\cite{Essig:2015cda, Dreyer:2023ovn, Knapen:2021run}, 
    \begin{equation}\label{eq:dielectric_cff}         
        |f_{\textrm{crystal}}(q, \Delta E)|^{2} = \frac{q^{5} V_\mathrm{cell}}{8 \pi^{2} m^{2} \alpha^{2}}\textrm{Im}\left({\epsilon_\mathrm{RPA}(q,\Delta E)}\right)\ ,
    \end{equation}
    where $\epsilon_\mathrm{RPA}(\Delta E, q)$ is the dielectric function calculated using the random phase approximation (RPA) \cite{Bohm:1951RPA1, Bohm:1952RPA2, Bohm:1953RPA3, Ehrenreich:1959RPA4}. We will use {\tt QEDark}~\cite{Essig:2015cda} and {\tt QCDark}~\cite{Dreyer:2023ovn} to compute the crystal form factor as additional numerical methods to obtain the material response. In order to accommodate the full response of the material, which requires also the real part of the dielectric function, we supplement our \texttt{QEDark}  and \texttt{QCDark}  calculations of the crystal form factor with the real part of the Lindhard dielectric function, Eq.~\eqref{eq:Lind}. We use the PBE functional~\cite{Perdew:1996PBE} with an $8\times8\times8\ \mathbf{k}-$grid for all calculations for \texttt{QEDark}. To obtain the \texttt{QCDark} derived form factor, we use a PBE functional~\cite{Perdew:1996PBE} for Si and PBE0~\cite{Adamo:1999PBE0} functional for Ge, both with a $6\times6\times6\ \mathbf{k}-$grid. 
 
\section{Results}\label{sec:results}

   We are now in a position to compute  Compton rates and compare our new approach that includes the dielectric material response to computations using RIA, {\tt FEFF}, and single-electron excitations. In what follows, we present results for Si, Ge, GaAs, and SiC (3C), but our calculations can easily be extended to other materials. 

   Fig.~\ref{fig:cs_rates} shows the Compton scattering rates for a 1 kg$\cdot$yr exposure in various crystals: Si ({\it top-left} panel), Ge ({\it top-right} panel), GaAs ({\it bottom-left} panel), and SiC (3C) ({\it bottom-right} panel), calculated using
    \begin{equation}
        \frac{dR}{d\Delta E} = n_\gamma\frac{M_\mathrm{Target}}{M_\mathrm{cell}}\frac{d\sigma_\omega}{d\Delta E}\ ,
    \end{equation}
    where $M_\mathrm{Target}$ and $M_\mathrm{cell}$ are the mass of the target material and the unit cell, respectively. We assume monochromic photons with energy $E_\gamma=1.461~\rm MeV$ and a density of $n_\gamma=1.25\times10^{-14}~\rm cm^{-3}$, which corresponds to a ionization background of 1~event/kg/day/keV (commonly called `DRU') at energies of 1~keV in Si.
    Our calculation predicts a factor of roughly 0.33 fewer events/kg/yr in Si for $Q \leq 5\ e^-$ compared to the prediction using RIA, and a much smoother spectrum. 

    For Si and SiC, we find that all methods including the analytical Lindhard model (see Appendix~\ref{app:Lindhard}) produce consistent results, except RIA. The consistency between methods that do not include all-electron effects, such as \texttt{DarkELF}, \texttt{QEDark}, and the Lindhard model, with methods that do include all--electron effects such as \texttt{EXCEED-DM} and \texttt{QCDark}, imply that all--electron effects are not important for Compton scattering.\footnote{All-electron effects are important for DM--electron scattering for electron recoil energies $\gtrsim$ 20~eV.  We discuss the difference with Compton scattering further in Appendix~\ref{app:Lindhard}.} 

    On the other hand, because the Lindhard approximation and \texttt{DarkELF} (GPAW) \cite{Knapen:2021run, Knapen:2021bwg} do not include information about the $3d$-shell in Ga, Ge, and As, they underestimate the Compton scattering cross section at higher recoil energies in Ge and GaAs target materials. Further, the usage of the hybrid PBE0 functional in the \texttt{QCDark} calculation for Ge affects the energy of the $3d$--shell compared to the PBE functional used in \texttt{QEDark} (for more details, see Ref.~\cite{Dreyer:2023ovn}). The energy obtained for the $3d$--shell using PBE0 lies between $28.6$~eV and 29.0~eV, which is in close agreement to the experimental values of $\sim 29.5$~eV~\cite{Bearden:1967}. 

    As previously discussed, the \texttt{QEDark} and \texttt{QCDark} curves in all panels of Fig.~\ref{fig:cs_rates} use the dielectric formalism, Eq.~\eqref{eq:res_die}, where we use Eq.~\eqref{eq:dielectric_cff} to calculate the imaginary part of the dielectric function, and use the real part of the Lindhard dielectric function, Eq.~\eqref{eq:Lind}. This enables one to encompass more accurately the many-body effects of the material response.  We show these effects in Fig.~\ref{fig:screening} for a Si crystal, where the dashed lines  include only the single-electron excitations that are captured by the imaginary part of the dielectric function, while the solid lines also include the real part of the Lindhard dielectric function to better capture the collective effects via the $1/|\epsilon(\Delta E, q)|^2$ factor. As we see, collective effects in crystals reduce the Compton scattering cross sections at low energies $\lesssim 25$~eV by about a factor of $\sim 3$.

\begin{figure}[t]
        \centering
        \includegraphics[width=\linewidth]{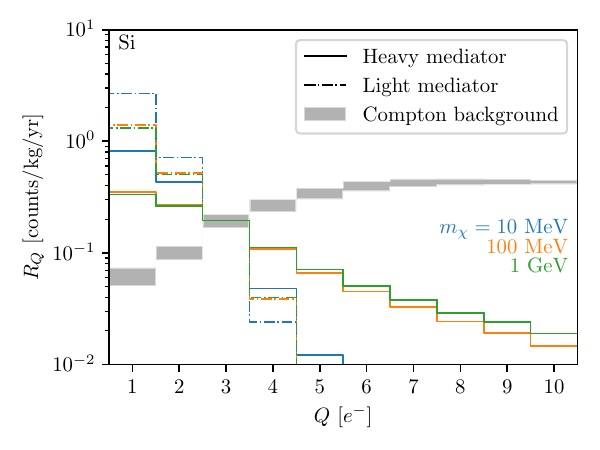}
        \caption{{\bf Rates as a function of charge for Si.} A comparison of the Compton scattering rates (\textit{gray} bands) as a function of the charge ionized, $Q$, in a Si crystal with a background rate of 1 DRU at $\Delta E = 1$ keV using RIA. The lines show the DM--electron scattering rates computed with \texttt{QCDark}~\cite{Dreyer:2023ovn} for DM--electron cross-sections listed in Table \ref{tab:dm_crosssections}. We use the secondary ionization model at 100~K from Ref.~\cite{ramanathan_ionization_2020}.}
        \label{fig:SiQ}
    \end{figure}

    Fig.~\ref{fig:SiQ} shows the expected Compton scattering rates for Si in terms of ionized charge following the secondary ionization modelling of Ref.~\cite{ramanathan_ionization_2020}. The \textit{gray} regions show the Compton scattering rates normalized to 1 DRU at $\Delta E = 1$ keV using RIA. The regions encompass the highest and lowest Compton scattering rates calculated using the Lindhard model, Eq.~\eqref{eq:Lind}, and using \texttt{DarkELF}~\cite{Knapen:2021bwg, Knapen:2021run}, \texttt{EXCEED-DM} \cite{Griffin:2021exdm, Trickle:2022exdm}, \texttt{QEDark} \cite{Essig:2015cda}, and \texttt{QCDark}~\cite{Dreyer:2023ovn}. The various colored lines show for comparison the DM--electron scattering rates for different DM masses $m_\chi$ computed using \texttt{QCDark}. We normalize the DM--electron scattering cross-sections to be equal to the Compton scattering cross-section in the $Q = 3\, e^-$ bin. Table~\ref{tab:dm_crosssections} lists the derived cross-sections. We note that the Compton spectrum decreases towards lower electron recoil energies, unlike the DM recoil spectra, which increase.  

    \begin{table}[]
        \centering
        \begin{tabular}{|c|c|c|}\hline
             \multirow{2}{*}{$m_\chi$}&\multicolumn{2}{c|}{$\bar{\sigma}_e$ [cm$^2$]} \\\cline{2-3}
             &Light mediator&Heavy mediator\\\hline\hline
             10 MeV& $2.20\times10^{-41}$ & $1.71\times10^{-42}$\\
             100 MeV& $8.41\times10^{-41}$ & $4.82\times10^{-42}$ \\
             1 GeV& $7.75\times10^{-40}$ & $4.42\times10^{-41}$\\\hline
        \end{tabular}
        \caption{{\bf Benchmark values.} DM mass and DM-electron scattering cross section benchmarks  for DM at reference momentum transfer $q = \alpha m_e$ used in the DM--electron scattering rates shown in Fig.~\ref{fig:SiQ}.}
        \label{tab:dm_crosssections}
    \end{table}
    
    \begin{figure}
        \centering
        \includegraphics[width=\linewidth]{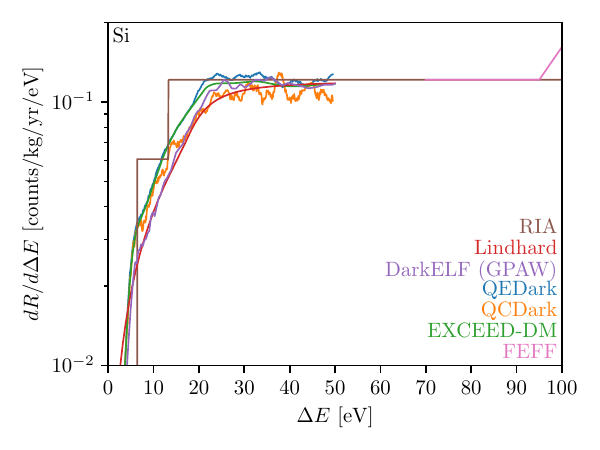}
        \caption{{\bf Spotlight on Si.} A reproduction of the {\it top-left} panel of Fig.~\ref{fig:cs_rates} for Si with the energy axis extended to 100~eV. We see that our results match onto those computed with RIA and {\tt FEFF} at high energies.}
        \label{fig:100eV}
    \end{figure}

    Fig.~\ref{fig:100eV} shows the expected Compton rates in Si for energies up to 100~eV, which shows how our results match onto those obtained using the \texttt{FEFF} code~\cite{Rehr:2000FEFF, Rehr:2009FEFF, Rehr:2010FEFF, Kas:2021}. The results from the {\tt FEFF} code were recently compared to experimental data by the DAMIC-M collaboration and found to be roughly consistent at energies $\Delta E \gtrsim \order{100\ \text{eV}}$~\cite{DAMIC-M:2022xtp}. 

\section{Summary}\label{sec:summary}
    In this paper, we have proposed a rigorous way to compute Compton scattering rates utilizing the material response of the target. This is particularly useful in the low-energy regime. We show how the dielectric function of any material can be used to precisely evaluate the Compton scattering cross section near the energy thresholds. The dielectric function approach is convenient as it incorporates the many-body effects of the system, including all-electron and band-structure effects.  The Compton scattering rate can then be calculated using existing codes, analytical forms, or experimental data. Importantly, we find that the simple analytical Lindhard function often captures well the behavior of the Compton scattering rate over a broad range of materials, allowing one to swiftly and accurately obtain the Compton rate. 
    
    As a concrete example, we calculated the event rate for Si, Ge, GaAs, and SiC with various methods. Compared with the conventional RIA method, our new approach, which includes band structure effects, leads to smoother Compton scattering cross-sections for energies below several 10's of~eV. It is necessary to include collective effects of the electron density, which lowers the Compton scattering rates for energies below $\sim 25$~eV. The all-electron corrections are found to be small for Compton scattering. 

\acknowledgments 

    We thank Cyrus Dreyer, Marivi Fern\'andez-Serra, Ryan Plestid, and Tanner Trickle for useful discussions. RE acknowledges support from DoE Grant DE-SC0009854, Simons Investigator in Physics Award 623940, and the US-Israel Binational Science Foundation Grant No.~2020220. The work of YH is supported by the Israel Science Foundation (grant No. 1818/22), by the Binational Science Foundation (grant No. 2018140) and by an ERC STG grant (`Light-Dark', grant No. 101040019). The work of YS is supported by the Israel Science Foundation (grant No. 1818/22). AS was supported by a Stony Brook IACS Seed Grant, from Fermilab subcontract 664693 for the DoE DMNI award for Oscura, from DoE Grant DE-SC0009854, and from the Simons Investigator in Physics Award 623940. GS was supported by DoE Grant DE-SC0009854 and Simons Investigator in Physics Award 623940. This project has received funding from the European Research Council (ERC) under the European Union’s Horizon Europe research and innovation programme (grant agreement No. 101040019).  Views and opinions expressed are however those of the author(s) only and do not necessarily reflect those of the European Union. The European Union cannot be held responsible for them.

\appendix

\section{Single-electron excitations in crystals}\label{app:relativistic}
    In general, the cross section for a $2 \to 2$ process is
    \begin{align}\label{eq:xsecgen}
	   \sigma_{\vec{p}, \vec{k}} &= \frac{1}{16 E_{p} E_{k}  v_{\text{rel}}} \int \frac{d^{3}q}{\left(2\pi\right)^{3}} \frac{d^{3}k'}{\left(2\pi\right)^{3}}\frac{ \overline{|{\cal M}|^{2}} }{E_{|\vec{p}-\vec{q}|} E_{k'}} \nonumber\\
		    &\times \left(2 \pi\right)^{4} \delta\left(E_i - E_f\right) \delta^{3}(\vec{k}+\vec{q}- \vec{k'})\ ,
    \end{align} 
    where the integration is over the final momenta of the two particles, $v_{\text{rel}}$ is the initial relative speed of the particles, and ${\cal M}$ is the amplitude that describes the interaction. For Compton scattering, $\vec{k}$ and $\vec{k'}$ are the initial and final momenta of the electron, and $\vec{p}$ and  $\vec{p'} = \vec{p}-\vec{q}$ are the initial and final momenta of the photon. $E_{i}$ and $E_{f}$ are the total initial and final energies of the system. A bound electron in the detector will have a non-relativistic momentum, so $v_{\text{rel}} \approx 1$.

    We now follow a similar approach to Ref.~\cite{Essig:2015cda}. The non-relativistic amplitude in free-electron Compton scattering is defined by
    \begin{align}\label{eq:free}
    	\bra{\gamma_{\vec{p}-\vec{q}}, e_{\vec{k'}}}&H_{\text{int}}\ket{\gamma_{\vec{p}},e_{\vec{k}}} \nonumber\\
    	&= C {\cal M}_{\text{free}}\left(2\pi\right)^{3} \delta^{3}(\vec{k}+\vec{q}-\vec{k'})\ ,
    \end{align} 
    where $C$ is an unimportant constant. For a bound electron in a detector, its state can be represented by a superposition of plane waves, $\ket{e} =  \sqrt{V} \int \frac{d^{3}k}{\left(2\pi\right)^{3}}\tilde{\psi}(\vec{k}) \ket{e_{\vec{k}}}$, so the overlap integral can be written as
     \begin{align}\label{eq:overlap}
    	&\bra{\gamma_{\vec{p}-\vec{q}}, e_{f}} H_{\text{int}}\ket{\gamma_{\vec{p}},e_{i}}\nonumber\\ 
    					  &\quad = V \int \frac{d^{3}k}{\left(2\pi\right)^{3}} \tilde{\psi}_{i}(\vec{k}) \tilde{\psi}_{f}^{*}(\vec{k}+\vec{q}) C {\cal M}_{\text{free}}\, .
    \end{align} 
    In general, ${\cal M}_{\text{free}}$ depends on the initial momentum of the electron and therefore, cannot be pulled out of the integral. However, assuming the electron is non-relativistic before and after the collision, which is consistent with energy transfers in the range $\lesssim$50~eV, our amplitude reduces to that of Thomson scattering~\cite{Peskin:1995ev}, 
    \begin{align}\label{eq:thompson}
    \overline{|{\cal M}_{\text{\rm free}}|^{2}} &= 2 e^{4} \left[ \frac{k \cdot p'}{ k\cdot p}+\frac{k\cdot p}{k \cdot p'}\right. \nonumber\\
    					 &+2m^{2}\left(\frac{1}{k\cdot p}-\frac{1}{k\cdot p'}\right)\nonumber \\
    					 &\left.+m^{4} \left(\frac{1}{k\cdot p}-\frac{1}{k\cdot p'} \right)^{2}  \right]\\
    					 &\approx 2e^{4}\left(1+\cos^{2}\theta \right)\ .
    \end{align} 
    The Thomson scattering amplitude does not depend on the initial momenta of the electron, so it can be factored out of the integral in Eq.~\eqref{eq:overlap}. By squaring Eqs.~\eqref{eq:free} and~\eqref{eq:overlap} and comparing them, we see that going from a free-electron to a bound electron amounts to the following substitution, 
    \begin{align}\label{eq:V}
    	\left(2\pi\right)^{3}&\overline{|{\cal M}_{\text{free}}|^{2}} V \delta^{3}(\vec{k}+\vec{q}-\vec{k'}) \nonumber\\
    \to & \quad V^{2} \overline{|{\cal M}_{\text{free}}|^{2}}|f_{i\to f}\left(\vec{q}\right)|^{2}\ ,
    \end{align} 
    where $f_{i\to f}\left(\vec{q}\right)$ is the atomic form factor and is defined as
     \begin{align}\label{eq:form}
    	f_{i \to f}\left(\vec{q}\right)=\int d^{3}x \psi_{i}\left(\vec{x}\right) \psi_{f}^{*}\left(\vec{x}\right) e^{i \vec{q}\cdot \vec{x}}.
    \end{align} 
    The factor of V on the left of Eq.~(\ref{eq:V}) is due to the redundant delta-function after squaring, which contributes a factor of $\frac{V}{\left(2\pi\right)^{3}}$. Applying this substitution in Eq.~(\ref{eq:xsecgen}), considering only a single final electron state, and letting $E$ and $E^\prime$ ($\omega$ and $\omega'$) be the initial and final energies of the electron (photon) yields
    \begin{align}\label{14}
    	\sigma_{\vec{p}, e_{i} \to  e_{f}} = \frac{1}{4 E \omega}& \int \frac{d^{3}q}{\left(2\pi\right)^{3}} \frac{1}{4 E' \omega'} \left(2\pi\right) \delta\left(E_i - E_f\right)\nonumber\\
    	&\times \overline{|{\cal M}_{\text{free}}|^{2}}|f_{i \to f}\left(\vec{q}\right)|^{2}\ .
    \end{align}
    
    The initial and final energies of the system are 
    \begin{align}\label{}
        E_{i} = \omega + m + E_{e, i}
    \end{align} 
    and
    \begin{align}\label{}
    	E_{f} &= \omega' + m + E_{e, f} \nonumber\\
        &=\sqrt{\omega^{2}+q^{2}-2 \omega q \cos\theta_{pq}}+m+E_{e,f}\ ,
    \end{align} 
    where $E_{e,i}$ and $E_{e,f}$ are the initial and final energies of the electron excluding the rest mass of the electron, i.e., $E_{e,i} = E - m$ and $E_{e,f} = E^\prime - m$. We can eliminate the delta-function by averaging over initial photon momenta directions. We assume a spherically symmetric distribution of incident photons, and we take $E, E^\prime \approx m$, since the electron is non-relativistic before and after scattering, which gives 
    \begin{align}\label{}
        \sigma_{\omega, e_{i} \to  e_{f}} = \frac{1}{128 \pi^{2} m^{2} \omega^{2}} \int d^{3}q \frac{1}{q}\overline{|{\cal M}_{\text{free}}|^{2}}|f_{i \to f}\left(\vec{q}\right)|^{2}\ .
    \end{align}
    We get the kinematic constraints $\Delta E \le  q \le 2\omega -  \Delta E
    $ and  $\Delta E \le  \omega$, where $\Delta E = E_{e,f} - E_{e,i}$. 
    
    For electrons in a periodic lattice, by Bloch's theorem, the electronic wavefunctions are described by a plane wave modulated by a periodic function,
    \begin{align}\label{eq:xsec}
        \psi_{i,\vec{k}}\left(\vec{x}\right) =\frac{1}{\sqrt{V} } \sum_{\vec{G}}^{} u_{i}(\vec{k}+\vec{G})e^{i (\vec{k}+\vec{G})\cdot \vec{x}}\ ,
    \end{align} 
    where $\vec{k}$ is a wavevector in the first Brillouin Zone, $\vec{G}$ is the reciprical lattice vector, and the subscript $i$ is the band index. The wavefunctions are normalized such that $\sum_{\vec{G}}^{} |u_{i}(\vec{k}+\vec{G})|^{2} = 1$.

    Substituting Eq.~(\ref{eq:xsec}) into Eq.~(\ref{eq:form}) and squaring yields
    \begin{align}\label{}
    	|f_{i,\vec{k}\to i',\vec{k'}}|^{2}&= \sum_{\vec{G'}}^{}\left(2\pi\right)^{3}\delta^{\left(3\right)}(\vec{q}-(\vec{k'}+\vec{G'}-\vec{k}))\nonumber\\
    	&\quad \quad\times \frac{1}{V}|f_{[i\vec{k},i' \vec{k'},\vec{G'}]}|^{2}\ ,
    \end{align} 
    where 
    \begin{align}\label{}
        f_{[i\vec{k},i' \vec{k'},\vec{G'}]} = \sum_{\vec{G}}^{}u^{*}_{i'}(\vec{k'}+\vec{G}+\vec{G'})u_{i}(\vec{k}+\vec{G}).
    \end{align} 
    We have replaced the subscripts $i\to f$ with the more explicit $i, \vec{k}\to i', \vec{k}'$. Then, our cross section becomes 
    \begin{align}\label{}
    	\sigma_{\omega, \{i, \vec{k}\} \to \{i', \vec{k'} \}} &= \frac{\pi}{16 m^{2}\omega^{2}} \frac{1}{q}\overline{|{\cal M}_{\text{free}}|^{2}}\nonumber\\
     &\times \sum_{\vec{G'}}^{} \frac{1}{V}|f_{[i\vec{k},i' \vec{k'},\vec{G'}]}|^{2}\biggr\rvert_{q = |\vec{k'}+\vec{G'}-\vec{k}|}\ .
    \end{align} 
    Summing over all initial and final electron states and multiplying by $2$ to account for the two degenerate spin states gives the total scattering cross section from the crystal, 
    \begin{align}
    	\sigma_{\omega} =  \frac{\pi}{8 m^{2}\omega^{2}}&V_{\text{cell}}N_{\text{cell}} \sum_{i i'}^{}\int_{\rm BZ} \frac{d^{3}k d^{3}k'}{\left(2\pi\right)^{6}} \frac{1}{q}\overline{|{\cal M}_{\text{free}}|^{2}}\nonumber\\
    	&\times  \sum_{\vec{G'}}^{}|f_{[i\vec{k},i' \vec{k'},\vec{G'}]}|^{2} \biggr\rvert_{q = |\vec{k'}+\vec{G'}-\vec{k}|}\ .
    \end{align}
    Here, $N_{\rm cell}$ and $V_{\rm cell}$ are the number of atoms in the primitive cell and the volume of the primitive cell, and the integral is over the first Brillouin zone in the crystal's reciprocal space. To get the differential cross section, we integrate over delta functions, leaving
    \begin{align}\label{}
        \frac{d\sigma_{\omega}}{d\left(\Delta E\right) dq} =  \frac{\alpha N_{\text{cell}}}{16 \pi \omega^{2}} \frac{1}{q^{2}}\overline{|{\cal M}_{\text{free}}|^{2}}|f_{\textrm{crystal}}\left(q,\Delta E\right)|^{2}\ ,
    \end{align} 
    where $f_{\textrm{crystal}}\left(q, \Delta E\right)$ is the crystal form factor~\cite{Essig:2015cda}, 
    \begin{align}\label{26}
    	&|f_{\textrm{crystal}}\left(q, \Delta E\right)|^{2} = \frac{2 \pi^{2}}{\alpha m^{2} V_{\text{cell}}} \sum_{i i'}^{}\int_{BZ} V_{\text{cell}}^{2} \frac{d^{3}k d^{3}k'}{\left(2\pi\right)^{6}}\nonumber\\
					      & \times \delta (\Delta E - E_{i',\vec{k'}}+E_{i,\vec{k}})\sum_{\vec{G'}}^{} q \delta(q-|\vec{k'}-\vec{k}+\vec{G'}|)\nonumber\\
					      &\times |f_{[i\vec{k},i' \vec{k'},\vec{G'}]}|^{2}\ .
    \end{align} 
    Lastly, we substitute in the Thomson scattering amplitude and divide by the number of cells to get the differential cross section per unit cell,
    \begin{align}\label{finalresult}
        \frac{d\sigma_{\omega}}{d\left(\Delta E\right) dq} =  \frac{2\pi \alpha^{3}}{\omega^{2}} \frac{\left(1+\cos^{2}\theta\right)}{q^{2}}|f_{\textrm{crystal}}\left(q,\Delta E\right)|^{2}\ .
    \end{align}

    Using Eq.~\eqref{eq:dielectric_cff}, we find that Eq.~\eqref{finalresult} can be written as
    \begin{equation}\label{eq:res_single_electron}
        \frac{\sigma_\omega}{d(d\Delta E)dq} = \frac{\alpha V_{\rm cell}}{4\pi m^2} \frac{q^{3}}{\omega^{2}}(1+\cos^{2}\theta)\textrm{Im}\left\{\epsilon(\Delta E, q)\right\}\ .
    \end{equation}
    Comparing to Eq.~\eqref{eq:res_die}, we see that the single-electron excitation approach misses a factor of $1/\abs{\epsilon(\Delta E, q)}^2$, and hence fails to capture the complete density response of the system as discussed in the main text. In effect, this approach misses the self-energy corrections to the photon in the target material. 
    
\section{Density Functional Theory}

    The dielectric function and the crystal form factor, as shown in Eqs.~(\ref{eq:res_die}) and (\ref{eq:dielectric_cff}), depend on the electronic structure of the target material. Here we provide a short review of Density Functional Theory (DFT), which is the typical method used to approximate the orbital wavefunctions and energy levels~\cite{Essig:2015cda, Knapen:2021run, Knapen:2021bwg, Griffin:2021exdm, Trickle:2022exdm}. DFT rests on two major theorems, known as the Hohenberg-Kohn Theorems \cite{Hohenberg:1964DFT}, 
    \begin{enumerate}
    	\item The energy of the ground state is a unique functional of the electron density.
    	\item The electron density that minimizes the energy is the correct solution to the ground state (Variational Theorem).
    \end{enumerate}
    
    The energy functional, in terms of the electron density $n(\vec{r})$, is given by~\cite{Kohn:1965DFT}
    \begin{equation}
        \begin{split}
            E[n] &= T_{\text{S}}[n] + \int\mathrm{d}^3 r v_{\text{ext}}(\vec{r})n(\vec{r}) \\ &+ 
            \frac{1}{2}\int\mathrm{d}^3 r\int\mathrm{d}^3 r^\prime\ \frac{n(\vec{r})n(\vec{r}^\prime)}{\abs{\vec{r} - \vec{r}^\prime}} + E_{\text{xc}}[n]\ ,
            \label{eq:KS}
        \end{split}
    \end{equation}
    where $T_\mathrm{S}$ is the kinetic energy of non-interacting orbitals, $v_\mathrm{ext}(\vec{r})$ is the external potential from the nuclei, and $E_\mathrm{xc}$, known as the exchange and correlation functional, models the quantum interactions missing in $T_\mathrm{S}$ and the Hartree electron--electron interaction (third term). 
    
    The minimization of the energy functional with the constraint that the total number of electrons remains constant leads to the Kohn--Sham equations~\cite{Kohn:1965DFT},
    \begin{align}\label{}
    	\left(-\frac{1}{2} \vec{\nabla}^{2} + \right. & v_{\text{ext}}\left(\vec{r}\right) + \int \left. \frac{n(\vec{r'})}{|\vec{r}-\vec{r'}|}d^{3}r' + v_\mathrm{xc}\left(\vec{r}\right) \right)\phi_{i}\left(\vec{r}\right)\nonumber \\
    	&= \epsilon_{i}\phi_{i}\left(\vec{r}\right)\ ,
    \end{align} 
    where $\epsilon_i$ and $\phi_i(\vec{r})$ are the energy levels and wavefunctions of the $i^\mathrm{th}$ orbital respectively, and 
    \begin{equation}
        v_\mathrm{xc}(\vec{r}) = \frac{\delta E_\mathrm{xc}[n]}{\delta n(\vec{r})}\ .
    \end{equation}
    Because the Hamiltonian depends on $n(\vec{r})$, which depends on the occupied orbitals $\left\{\phi_i(\vec{r})\right\}$, this equation must be solved self--consistently.
    
\begin{figure*}
    \centering
    \includegraphics[width =0.497\linewidth]{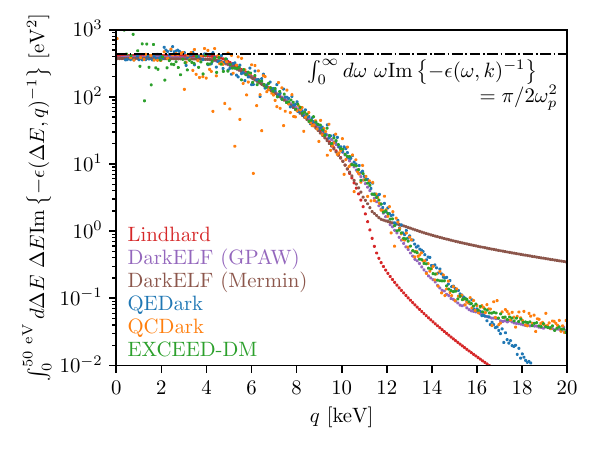}
    \includegraphics[width =0.497\linewidth]{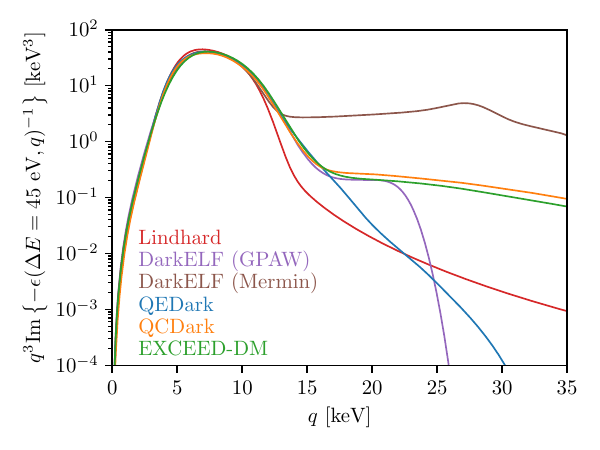}
    \caption{\textit{Left panel:} The $f-$sum rule~\cite{Bassani:2005sumrule} for the loss function calculated for Si using various methods. \textit{Right panel:} The electron loss function $\mathrm{Im}\left\{-\epsilon(\Delta E, q)^{-1}\right\}$ for Si, weighted by momenta factors relevant for Compton scattering, calculated at $\Delta E = 45$ eV. For this plot, we average the electron loss function over a 2 eV bin centered at 45~eV, and apply a 1D Gaussian filter in $k$ with $\sigma_k = 1$~keV. Note the presence of high-$q$ modes due to all-electron modelling in \texttt{QCDark} and \texttt{EXCEED-DM} compared to \texttt{DarkELF} and \texttt{QEDark}.}
    \label{fig:epsilon}
\end{figure*}

    The exact form of the exchange and correlation functional is not known, though there are multiple models available. In this paper, we use the PBE functional~\cite{Perdew:1996PBE} for both \texttt{QEDark} and \texttt{QCDark} in Si. For Ge, we use PBE and PBE0~\cite{Adamo:1999PBE0} functionals in \texttt{QEDark} and \texttt{QCDark}, respectively. We use the standard available results for  \texttt{DarkELF}~\cite{Knapen:2021bwg} and the {\it valence-to-conduction-band-only} results for \texttt{EXCEED-DM}~\cite{Trickle:2022exdm}. 

\section{Effectiveness of the Lindhard Model}\label{app:Lindhard}

    The Lindhard model produces reliable results for the Compton scattering cross section for valence electrons, and hence produces reliable results until the effects of core electrons become important. For example, for Si, the Lindhard model produces reliable results for $\Delta E \lesssim 99$~eV, while for Ge it produces reliable results for $\Delta E \lesssim 29.2$~eV. 
    
    The left panel of Fig.~\ref{fig:epsilon} shows the $f-$sum rule~\cite{Bassani:2005sumrule} for the loss function calculated for Si using various methods. We see that the Mermin model overestimates the electron loss function, $\mathrm{Im}\left\{-1/\epsilon(\Delta E, q)\right\}$, for $q \gtrsim 12$~keV, and further note that the Lindhard model underestimates the loss function at high $q \gtrsim 12$ keV. 
    
    The differential cross section for any general process, where an incoming particle with speed $\beta$ relative to the target material transfers energy $\Delta E$ and momentum $q$ to an electron, is
    \begin{equation}
        \dv{\sigma}{\Delta E} = \int_{q_\mathrm{min}}^\infty~dq~\pdv{\sigma}{\Delta E}{q}\ ,
    \end{equation}
    where $q_\mathrm{min} = \Delta E/\beta$ is the minimum momentum transfer required to transfer energy $\Delta E$. Because in the case of Compton scattering, the incoming particles are photons, $\beta = 1$, and so integrating Eq.~\eqref{eq:res_die} in $q$, one obtains
    \begin{equation}
        \frac{d\sigma_\omega}{d\Delta E} = \frac{\alpha V_{\rm cell}}{4\pi m^2} \int_{\Delta E}^\infty dq\ \frac{q^{3}}{\omega^{2}}(1+\cos^{2}\theta)\textrm{Im}\left(\frac{-1}{\epsilon(\Delta E, q)}\right)\ .
    \end{equation}
    Hence, the peak in the electron loss function $\mathrm{Im}\left\{-\epsilon(\Delta E, q)^{-1}\right\}$ dominates. This is demonstrated in the right panel of Fig.~\ref{fig:epsilon}. The low-$q$ region is captured well by the Lindhard model for valence electrons, and hence the Lindhard model provides reliable Compton scattering rates in the low energy regime. 
    
    This is in contrast to DM--electron scattering where the incoming DM particles have a speed $\beta \sim \order{10^{-3}}$, and so for $\Delta E \sim \order{10\ \mathrm{eV}}$, the lowest possible momentum transfer is $q_\mathrm{min} \sim \order{10\ \mathrm{keV}}$. Further, capturing the electron loss function at high momentum transfer makes inclusion of all--electron effects  necessary \cite{Griffin:2021exdm, Trickle:2022exdm, Dreyer:2023ovn}.
    The Lindhard response function of the material thus captures the Compton scattering rates well, but does not capture well the DM-electron scattering rates~\cite{Hochberg:2021pkt,Knapen:2021run}. 
    
\newpage
\bibliography{references}

\end{document}